\documentclass[a4paper,reqno]{amsart}
\usepackage{setspace}
\usepackage{hyperref}
\usepackage{amsmath}
\usepackage{amsthm}
\usepackage{amssymb}
\usepackage{amsfonts}
\usepackage{mathrsfs}
\usepackage{mathtools}
\usepackage{graphicx} 
\usepackage{cite}
\usepackage{hyperref} 

\usepackage{chngpage}

\usepackage{tablefootnote}

\numberwithin{equation}{section}

\newcommand\weak{\doteq}

\newcommand\bfE{\mathbf{E}}
\newcommand\bfB{\mathbf{B}}

\newcommand{\tb}[1]{\textbf{#1}}

\usepackage{color}

\newcounter{mnotecount}[section]

\theoremstyle{plain}
\newtheorem{theorem}{Theorem}[section]

\newtheorem{remark}[theorem]{Remark}

\newcommand{\pd}{\partial}

\newcommand\rd{\bar\pd} 
\newcommand\pr{\text{\textsf{pr}}} 

\newcommand{\dual}[1]{{}^\star\!#1}

\newcommand{\dualF}{\dual{F}}

\newcommand\Ascr{{\mathcal A}} 
\newcommand\Escr{{\mathcal E}}
\newcommand\Fscr{{\mathcal F}} 
\newcommand\Lscr{{\mathcal L}}
\newcommand\Mscr{{\mathcal M}} 
 
 \newcommand\Uscr{{\mathcal U}}
\newcommand\Xscr{{\mathcal X}} \newcommand\Zscr{{\mathcal Z}}

\newcommand\barAscr{\hbox{$\bar{\phantom{\Ascr}}\!\!\!\!\!\Ascr$}}
\newcommand\barFscr{{\bar{\Fscr}}}

\newcommand\Esf{\mathsf{E}} 
\newcommand\unit{\text{I}} 

\newcommand\cc{\text{c.c.}} 

\renewcommand{\div}{\mathop{\rm div}\nolimits}
\newcommand\tr{\mathop{\rm tr}\nolimits}

\newcommand{\halfspace}{\kern.083333em}   
\newcommand{\quart}{\kern.0416675em}  
\newcommand{\nhalf}{\kern-.083333em}   
\newcommand{\nquart}{\kern-.0416675em}  

\title[Zilch conservation revisited]{The zilch electromagnetic conservation law revisited
}

\author[S. Aghapour]{Sajad Aghapour}
\address[S. Aghapour]{School of Physics, Institute for Research in Fundamental Sciences (IPM), Tehran, Iran \and Albert Einstein Institute, Am M\"uhlenberg 1, D-14476 Potsdam, Germany}
\email{aghapour@ipm.ir}

\author[L. Andersson]{Lars Andersson}
\address[L. Andersson]{Albert Einstein Institute, Am M\"uhlenberg 1, D-14476 Potsdam, Germany }
\email{laan@aei.mpg.de}

\author[K. Rosquist]{Kjell Rosquist}
\address[K. Rosquist]{AlbaNova University Center, Stockholm University, Department of Physics, SE-106 91 Stockholm, Sweden}
\email{kr@fysik.su.se}

\begin{document}

\date{\today}

\maketitle


\begin{abstract}
It is shown that the zilch conservation law arises as the Noether current corresponding to a variational symmetry of a duality-symmetric Maxwell Lagrangian. The action of the corresponding symmetry generator on the duality-symmetric Lagrangian, while non-vanishing, is a total divergence as required by the Noether theory. The variational nature of the zilch conservation law was previously known only for some of the components of the zilch tensor, notably the optical chirality. By contrast, our analysis is fully covariant and is therefore valid for all components of the zilch tensor. The analysis is presented here for both real and complex versions of duality-symmetric Maxwell Lagrangians.
\end{abstract}


\section{Introduction}
Conservation laws of Maxwell's electromagnetic field equations, a subject  with a history going back more than a century, has in the last decades received renewed attention driven by fundamental developments in both theory and experiment. In recent studies, a new set of conserved quantities have been found, the most remarkable ones being the optical helicity, optical spin and orbital angular momenta and optical chirality of electromagnetic fields. The concept of electromagnetic or optical helicity was introduced by Trueba and Ra{\~{n}}ada based on similar notions previously defined in magnetohydrodynamics and fluid dynamics \cite{Trueba_1996}. However, a major progress in understanding the optical helicity as a conserved quantity occurred only in recent years (see for example \cite{Cameron2012a} and references therein). These concepts play an important role in the interaction of light beams with chiral molecules and nanoparticles. Besides its definition in free space and linear media, the optical helicity has also been introduced for fields in dispersive media \cite{PhysRevLett.120.243605}. The conservation laws for spin and orbital parts of the angular momentum, previously introduced by Allen et al.\ \cite{1992PhRvA..45.8185A} and van Enk and Nijenhuis \cite{1994JMOp...41..963V}, are also considered and reexamined recently by several groups, including I.\ and Z.\  Bialynicki-Birula  \cite{Bialynicki-Birula2011}, Cameron and Barnett \cite{Cameron&Barnett:2012}, and Bliokh et al. \cite{Bliokh_2014}. A comprehensive review on different types of optical angular momenta and their properties is given in   \cite{BLIOKH20151}. The angular momentum of monochromatic light in complex media is studied in \cite{Bliokh_2017}. Remarkably, some of these notions also extend to other physical fields such as acoustic waves and linearized gravity fields, cf. e.g. \cite{PhysRevB.99.020301,PhysRevB.99.174310,PhysRevLett.123.054301,PhysRevLett.123.183901,Burns_2020,Barnett2014,Andersson_etal:2018}.

Optical chirality is another related quantity in this context with a fairly long history that has attracted new attention recently. It was noted by Lipkin in his 1964 paper \cite{Lipkin:1964} that the pseudoscalar $C = \bfE \cdot \nabla\times \bfE + \bfB \cdot \nabla\times\bfB$, now referred to as the optical chirality \cite{Tang&Cohen:2010}, satisfies a differential conservation law. He also showed, referring to an observer 4-velocity $u^a = (1,0,0,0)$, that $C = Z^{000} = -Z_{abc} u^a u^b u^c$ is the component  of a conserved Lorentz covariant tensor named the zilch tensor $Z_{abc}$.\footnote{Cartesian coordinates will be used throughout.}
This tensor represents a collection of conserved currents parametrized by its first two indices. The pseudoscalar nature of $C$ makes it suitable for use in experimental investigations involving chiral electromagnetic fields such as circularly polarized light \cite{Tang&Cohen:2010}. The relation of the chirality of monochromatic optical fields to their helicity, spin angular momentum, and energy density has been studied by Bliokh and Nori \cite{PhysRevA.83.021803}, and its measurement with a plasmonic slit is reported in \cite{PhysRevLett.109.013901}. In \cite{Smith&Strange:2018}, the authors have suggested that the zilch tensor can encode information about the topology of electromagnetic field lines. Generation of a chiral current in a rotating photon gas with duality-violating boundary conditions is studied in \cite{Chernodub:2018}, where the temperature dependence is compared with that of a similar effect known for chiral fermions.

While the zilch tensor has been known for several decades as a conserved quantity, our understanding of its relation  to a symmetry of the Maxwell action still contains some gaps. In particular, a symmetry generator which leaves the standard Maxwell action invariant and which has the zilch as its Noether current has not been given in the literature. In earlier work, authors have considered symmetries leading to conservation of the scalar part $C$ (Deser and Teitelboim \cite{Deser&Teitelboim:1976} and Philbin \cite{Philbin:2013}) and of a vector part (Calkin \cite{Calkin:1965}) of the zilch tensor. Przanowski et al.~ \cite{Przanowski_etal:1994} related the first order action of \cite{Deser&Teitelboim:1976} to the zilch by a third order symmetry generator. Both Morgan and Joseph \cite{Morgan&Joseph:1965} and Sudbery \cite{Sudbery:1986} use a vector Lagrangian which is just the nonzero trace of the zilch tensor. Using the field tensor as the independent variable, it gives field equations which are equivalent to the Maxwell equations. They showed that the Lagrangian constructed in that way is related to the zilch tensor by spacetime translation symmetries.

As is well-known, Noether proved that a symmetry of an action leads to a conservation law. She also proved the less frequently used converse that a conservation law for some Euler-Lagrange equations emanates from a symmetry of the corresponding Lagrangian density. We refer to Olver's book for a comprehensive account of Noether's theorems \cite[see especially Theorem 5.58]{Olver:1993}. This equivalence between symmetries and conservation laws requires the consideration of \emph{generalized symmetries}, the generators of which depend not only on the dependent variables themselves but also on their derivatives up to some (finite) order (see \eqref{v_vector} below). In this work we have adopted the jet space formalism for analyzing symmetries \cite{Olver:1993}. A short introduction of the formalism is given in
Sect.~\ref{jet_intro}.

The main focus in this work is a clarification of the variational nature 
of the zilch conservation law by writing it in \emph{characteristic form} \cite{Olver:1993},
symbolically as $\div \!Z = Q \cdot \Delta$, and in full as
\begin{equation}\label{charform}
   \frac{\pd Z_{ab}{}^c}{\pd x^c} = Q_{abc} \Delta^c
\end{equation}
where the \emph{characteristic} $Q_{abc}$ represents the $ab$-parametrized components of the symmetry generator and $\Delta^a = \Esf^a(L)$ are the Euler-Lagrange expressions for a Lagrangian $L(A)$ as a function of the vector potential. The variation of the Lagrangian under the symmetry generator $Q_{abc} \,\pd/\pd A_c$ must then be zero modulo a total divergence \cite[Sect.~5.3]{Olver:1993}. 

The Maxwell equations are invariant under the 1-dimensional group of duality transformations (see e.g.~ \cite{Andersson_etal:2018})
\begin{equation}\label{duality}
 \begin{split}
        F_{ab} &\rightarrow
  \phantom{-}F_{ab} \halfspace\cos\alpha + \dualF_{ab}
       											 \halfspace\sin\alpha \\[1pt]
   \dualF_{ab} &\rightarrow
            -F_{ab} \halfspace\sin\alpha + \dualF_{ab} \halfspace\cos\alpha
 \end{split}
\end{equation}
It can be shown that the duality transformation \eqref{duality} leads to the conservation of the physically significant helicity current, cf. \cite{Deser&Teitelboim:1976,Cameron&Barnett:2012,Bliokh_etal:2013}, see also \cite{Andersson_etal:2018} and references therein. 
The original definition of the zilch tensor by Lipkin \cite{Lipkin:1964} and Kibble \cite{Kibble:1965} was formulated in the standard Maxwell framework without consideration of duality symmetry. However, when considering its relation to the conservation of helicity, and in particular to its relation to the helicity array introduced by Cameron et al. \cite{Cameron2012a}, a duality-symmetric formulation is necessary. 

To express the duality symmetry, the Maxwell theory is extended to include a magnetic 4-potential $C_a$ for the dual field tensor $\dual{F}_{ab}$ as an independent variable in addition to the electric 4-potential $A_a$, cf. \cite{Cameron_2013} and references therein. It is then possible to define a duality-symmetric Lagrangian for the extended theory as is done in Sect.~\ref{realdual} below. The Maxwell theory can also be formulated in a more compact way by using a complex potential $\Ascr_a = A_a + i C_a$ and a corresponding self-dual complex field tensor $\Fscr_{ab} = \frac12 (F_{ab} + i \dual{F})_{ab}$ \cite{Bliokh_etal:2013}. As an example, the duality transformations of the complex potential and the self-dual field tensor take the simple forms $\Ascr_a \rightarrow e^{-i\alpha} \Ascr_a$ and $\Fscr_{ab} \rightarrow e^{-i\alpha} \Fscr_{ab}$. This formulation has the additional advantage of making the duality symmetry manifest in the Lagrangian \cite{Bliokh_etal:2013,DRESSEL20151}, here given in \eqref{ds_lag}.
When studying the variational aspects of the zilch symmetry we will discuss two alternative duality-symmetric Lagrangians, one real and one complex described in Sect.\ref{dualinv}.  Since the two formulations are essentially equivalent we treat the real formulation in detail and give a shortened version of the complex formulation for completeness. 

Cameron and Barnett \cite{Cameron&Barnett:2012} discussed the variation of such a duality-symmetric Lagrangian with respect to a transformation which in our formalism corresponds to characteristics of the form $Q_{abc} = F_{c{(a,b)}}$. In doing so, they showed that the Lagrangian was left invariant, but only modulo the Euler-Lagrange equations (see \cite[Sect.4]{Cameron&Barnett:2012}). Therefore their analysis did not in fact establish the variational character of the transformation according to the Noether theory as described above. Nevertheless, we will show that the transformation they discussed is a variational symmetry as defined in \cite[Definition 5.51]{Olver:1993}. In addition to the works mentioned above, a number of authors have discussed and classified symmetries and conservation laws of the Maxwell equations without considering possible relations to a variational formulation, notably Fushchich and Nikitin \cite{Fuschchich&Nikitin:1983}, Pohjanpelto \cite{Pohjanpelto:1995}, Anco and Pohjanpelto \cite{Anco&Pohjanpelto:2001}. 

An important aspect of the characteristic form relation \eqref{charform} is that it is an identity. In this work we will be careful to make the important  distinction in the Noether theory between equations which are only valid modulo solutions of the field equations (then said to be valid \emph{on-shell}) and those that are identities (then said to be valid \emph{off-shell}). For example, equation \eqref{charform} implies the on-shell equation
\begin{equation}
   \frac{\pd Z_{ab}{}^c}{\pd x^c} \weak 0 \ ,
\end{equation}
where the notation ``$\halfspace\weak\halfspace$" signifies the on-shell restriction of the equality.
When dealing with symmetries and conservation laws, a common situation is that a given Lagrangian has a known symmetry and one can then derive a corresponding conservation law more or less straightforwardly using Noether theory. It may happen though, as is the case with the zilch tensor, that a conservation law is known but not the corresponding Noether symmetry. In this latter case one can look for a relation of the type \eqref{charform} and in that way identify the corresponding variational symmetry.
However, finding the symmetry using this reverse Noether procedure is not algorithmic and can be quite difficult. The underlying reason is that neither the symmetry nor the conservation law is uniquely defined. Rather they are only determined up to equivalence (see section \ref{jet_intro} for more details). 

\subsection*{Overview of this paper} In section \ref{jet_intro}, we briefly review the jet space formalism for analyzing the generalized symmetries of a field theory. It is our main tool for a careful application of the reverse Noether theorem to a conservation law in order to derive the associated symmetry of the variational problem. In section \ref{sec:zilch}, we present various forms of the Lipkin's zilch tensor of electromagnetic fields and its conservation law. The duality symmetric formulation of the Maxwell theory is discussed in \ref{dualinv}. Equipped with these tools, we study the conservation law of the zilch tensor in real duality-symmetric formulation of Maxwell theory in section \ref{realdual} and derive the variational symmetry underlying this conservation law. We complement this study with a review of its complex duality-symmetric formulation in section \ref{complexdual}. A summary of results are written in table \ref{table} and concluding remarks are given in section \ref{sec:concl}. For completeness, the equivalent forms of the zilch tensor are addressed in appendix \ref{app:alternative_forms} and its $1+3$ decomposition is given in appendix \ref{app:decomposition}.

\section{The jet space formalism for analyzing symmetries}\label{jet_intro}
The mathematical technique for analyzing symmetries adopted in this note is the jet space formalism, cf. Olver \cite{Olver:1993}, in which the infinitesimal action of a symmetry is represented by a linear differential operator. Such symmetry operators are then in one-to-one correspondence to a subspace of vector fields on a (finite-dimensional) jet space. In particular, they form a Lie algebra and the symmetry structure can in principle be understood by exploring that algebra without the need for a Hamiltonian formulation.
For Maxwell theory,\footnote{See also \cite{Pohjanpelto:1995} for a jet space tutorial adapted to the Maxwell equations and \cite{Rosquist:1989} for an application in general relativity, using tensor formalism.} the basic $n$-jet space has the form $M^{(n)} = M \times U_n$ where $M = (x^a, A_a)$ consists of the spacetime coordinates and the components of the 4-potential while each $U_k$ $(k\leq n)$ consists of the components of the $k$'th order derivatives of $A_a$. The elements of $M^{(n)}$ are to be considered as independent in calculations. By including derivatives of any finite order, this formalism is not restricted to point symmetries but can also handle generalized symmetries as needed for the zilch tensor for example.
In general, a symmetry generator has the form
\begin{equation}\label{v_vector}
   \xi^a \frac{\rd}{\rd x^a} + \phi_a \frac{\pd}{\pd A_a}
\end{equation}
where $\xi^a$ and $\phi_a$ are functions on $M$ for point symmetries and on some jet space $M^{(n)}$ for generalized symmetries. The bars on the first derivative in \eqref{v_vector} serves to indicate that $\rd/\rd x^a$ is restricted to act only on explicit occurrences of $x^a$.\footnote{For reference we note that Olver in \cite{Olver:1993} uses an opposite convention and instead refers to the standard partial coordinate derivative as a ``total derivative" $D_a$. Specifically, the relation to our notation is $(\pd/\pd x^a,\rd/\rd x^a) \leftrightarrow (D_a,\pd/\pd x^a)$. Our use of the comma notation will always stand for the standard partial derivative, as in $A_{a,b} = \pd A_a/\pd x^b$ for example.} For a point symmetry, $v$ is a vector field on $M$ which generates a Lie group of transformations. 

Any symmetry can be represented by a single field $Q_a = \phi_a -A_{a,b} \halfspace \xi^b $ with the corresponding generator
\begin{equation}\label{vQ}
   v = Q_a \frac{\pd}{\pd A_a}
\end{equation}
where the characteristic $Q_a$ is a jet space function. It could depend on the coordinates $x^a$, $A_a$ and derivatives of $A_a$ up to a finite order.
For the action of a symmetry on the Lagrangian or the Euler-Lagrange equations one needs to prolong the symmetry \eqref{vQ} (or \eqref{v_vector}) to take into account its action also on the derivatives of $A_a$. For functions on a jet space $M^{(n)}$, that prolongation is given by
\begin{equation}
   \pr^{(n)}v = v + \sum_{k=1}^n Q_{a,b_1\ldots b_k}\,
   					    \frac{\pd}{\pd A_{a,b_1\ldots b_k}}	\ .
\end{equation}
The condition for $v$ to be a variational symmetry for a Lagrangian $L$ is
\begin{equation}\label{varsymm}
   \pr^{(n)}v (L) = U^a{}_{,a}
\end{equation}
for some jet space field $U^a$ \cite[Ch.5]{Olver:1993}. We emphasize that \eqref{varsymm} must be an identity, i.e.~ be valid off-shell. The corresponding conserved current is then given by
\begin{equation}
   J^a = -Q_b \frac{\pd L}{\pd A_{b,a}} + U^a \, .
\end{equation}

In general, if a conserved current is a variational symmetry for a Lagrangian $L$, the conservation law takes the form
\begin{equation}\label{conslaw1}
   J^a{}_{,a} = Q_a \Escr^a
\end{equation}
where the characteristic $Q_a$ and $\Escr^a \weak 0$ are jet space functions.  The simplest case occurs when the relation has the characteristic form
\begin{equation}\label{conslaw2}
   J^a{}_{,a} = Q_a \halfspace\Esf^a(L) \, .
\end{equation}
This form of the conservation law therefore neatly expresses the two-way  nature of Noether's theorem:
\begin{equation*}
   \text{variational symmetry}
     \Leftrightarrow \text{conservation law}
\end{equation*}

For currents depending on higher derivatives of the dependent variables, such as the zilch tensor, the conservation law corresponding to \eqref{conslaw1} can also include derivatives of the field equations.
In that case one can recover the corresponding symmetry generator  
by performing partial integrations leading to a modified current $\tilde J^a$ obeying \eqref{conslaw2}. The modified current will then differ from the original by a part $K^a = \tilde J^a - J^a$, which vanishes on-shell, $K^a \weak 0$. The difference $K^a$ itself is said to be a trivial conservation law of the first kind and the conservation laws $J$ and $\tilde J$ are then said to be equivalent. We remark that for currents, there is a second notion of triviality. The difference between two currents is said to be trivial of the second kind if its divergence is identically zero regardless of any field equations. This shall be relevant in section \ref{realdual} below. It is worth noting that in general, the characteristic form \eqref{conslaw2} of a conservation law is not preserved if the conserved current is replaced by an equivalent one. The procedure described above will be carried out explicitly for the zilch tensor in Sect.\ \ref{dualinv} where the characteristic form for the zilch conservation law of the type \eqref{conslaw2} will be established. 


\section{The zilch tensor} \label{sec:zilch}
In this section we display for reference the basic algebraic and differential properties of the zilch tensor. Its original covariant form given by Lipkin was subsequently simplified by Morgan \cite{Morgan:1964} and Kibble \cite{Kibble:1965} to the form
\begin{equation}\label{ZKibble}
  Z_{abc} = \dualF_{ad} F^d{}_{b,c}-F_{ad}\halfspace\dualF^d{}_{b,c} \ .
\end{equation}
Although not apparent from this form, the zilch tensor is symmetric and traceless with respect to its first two indices
\begin{equation}\label{Zsym}
   Z_{abc} = Z_{(ab)c} \ ,\quad  Z^a{}_{ac} = 0 \ .
\end{equation}
To show this, it is convenient to express \eqref{ZKibble} in the (vector valued) matrix form as
\begin{equation}\label{Zc}
   Z_c = \dualF \halfspace F_{,c} - F \halfspace\dualF_{,c}
\end{equation}
where $F$ has components $F^a{}_b$. Following Kibble \cite{Kibble:1965} we use the following identity (valid for any antisymmetric matrix) to verify \eqref{Zsym}
\begin{equation}\label{Fid}
   F \halfspace\dualF = \tfrac14 \tr(F \halfspace\dualF) \ .
\end{equation}
Differentiating this expression gives
\begin{equation}
   F' \halfspace\dualF + F\halfspace \dualF' = \tfrac12 \tr(\dualF F') \ .
\end{equation}
We can then write the zilch tensor in the form
\begin{equation}
   Z_c = \dualF F_{,c} + F_{,c}\dualF - \tfrac12 \tr(\dualF F_{,c})
       = \{\dualF , F_{,c}\} - \tfrac12 \tr(\dualF F_{,c})
\end{equation}
using the anticommutator\footnote{\label{foot:anti} For any two matrices $H$ and $K$ the anticommutator is defined by $\{H,K\}= HK+KH$.}  notation in the last equality. 
In components this expression reads
\begin{equation}
   Z_{abc} = 2\halfspace\dualF^d{}_{(a} F_{b)d,c}
               -\tfrac12 g_{ab} \,\dualF^d{}_e \halfspace F^e{}_{d,c}
\end{equation}
in which the symmetry of its first two indices is manifest. The traceless property with respect to its first two indices also follows directly.

To show the validity of the conservation law $Z_{ab}{}^c{}_{,c} \weak 0$, Kibble used a third form of the zilch tensor which follows from \eqref{ZKibble} and \eqref{Zsym} and is given by\footnote{Kibble \cite{Kibble:1965} expressed this form in a different somewhat non-standard notation.} \cite{Kibble:1965}
\begin{equation}\label{zilchlr}
   Z_{abc} = \dualF^d{}_{(a} F_{b)d,c} - F^d{}_{(a} \dualF_{b)d,c} \ .
\end{equation}
In matrix notation this corresponds to
\begin{equation}\label{Z3}
   Z_c = \tfrac12 \{ \dualF, \pd_c F \} - \tfrac12 \{ F, \pd_c \dualF \}  \,.
\end{equation}
Taking the divergence gives
\begin{equation}\label{lrdiv}
  Z^c{}_{,c} = \tfrac12 \{ \dualF, \square F \}
             - \tfrac12 \{ F, \square \dualF \} \,.
\end{equation}
where the box operator stands for the d'Alembertian $\square = g^{ab}\pd_a \pd_b$. Expressing this divergence in components, the zilch conservation law then reads
\begin{equation}
   Z_{ab}{}^c{}_{,c} = \dualF^d{}_{(a} \,\square F_{b)d}
                      -F^d{}_{(a} \,\square \halfspace\dualF_{b)d}
                     \weak 0 \ .
\end{equation}
The conclusion that the divergence vanishes on solutions of the Maxwell equations follows from the fact that both $\square F_{ab}$ and $\square \dualF_{ab}$ vanish on-shell.

\section{Duality-symmetric formulation of Maxwell theory}\label{dualinv}
Here, we discuss two versions of duality-symmetric Lagrangians for the Maxwell theory, one real and one complex. In the duality-symmetric theory, the degrees of freedom are doubled. It becomes equivalent to standard Maxwell theory only after imposing a duality-constraint, see below. 
Our aim is to find the symmetry associated to the conservation of the zilch tensor and showing that it is a variational symmetry, i.e. the action which gives rise to the Maxwell equations is invariant under that symmetry. Here, we show this for duality-symmetric Lagrangian. We will further demonstrate that the zilch tensor is equivalent to the Noether current associated to this variational symmetry applied to the duality-symmetric Lagrangian.

The standard Maxwell Lagrangian is given by
\begin{equation}\label{sLag}
   L = \tfrac14 \tr (F^2) = \tfrac14 F^a{}_b F^b{}_a 
     = -\kappa^{abcd} A_{a,b} A_{c,d}
\end{equation}
where $F_{ab} =-2A_{[a,b]}$ is the field tensor expressed in terms of the 4-potential $A_a$ and $\kappa_{abcd}= g_{c[a} g_{b]d}$ is the antisymmetry projector.\footnote{This tensor projects out the antisymmetric part of any index pair, e.g.\ $M_{[ab]} = \kappa_{ab}{}^{cd} M_{cd}$. It has the same index symmetries as the Riemann tensor, $\kappa_{abcd} = \kappa_{cdab} = \kappa_{[ab]cd}$ and $\kappa_{a[bcd]} =0$.} The Euler-Lagrange equations then become
\begin{equation}
   \Esf^a(L) = -F^{ab}{}_{,b} = 0
\end{equation}
where $\Esf^a$ is the (first order) Euler operator
\begin{equation}
   \Esf^a = \frac{\pd}{\pd A_a} - \pd_b \frac{\pd}{\pd A_{a,b}}
\end{equation} 

The zilch tensor is invariant under the duality transformation \eqref{duality} but the standard Maxwell Lagrangian \eqref{sLag} is not.
To treat the zilch conservation law, we will consider two versions of duality-symmetric Lagrangians, one using a two-potential real formalism and one using complex variables, both of these introduced in \cite{Bliokh_etal:2013}. The complex field tensor is defined by
\begin{equation}
   \Fscr_{ab} = \tfrac12(F_{ab}+ i\dualF_{ab}) \ .
\end{equation}
The duality transformation now takes the simple form
\begin{equation}
   \Fscr_{ab} \rightarrow e^{-i\alpha} \halfspace\Fscr_{ab} \,.
\end{equation}
To go from the standard Lagrangian to one which is duality-symmetric we can start from the relations
\begin{equation}\label{FF}
   F^2 = (\Fscr + \overline\Fscr)^2 = X_1 + X_2
\end{equation}
where
\begin{align}
    X_1 &= \Fscr^2 + \bar\Fscr^2 = \tfrac12 [F^2 - (\dualF)^2] \\
    X_2 &= 2\Fscr \bar\Fscr =  \tfrac12 [F^2 + (\dualF)^2]
\end{align}
using the fact that $\Fscr$ and $\barFscr$ commute which in turn follows from the following identity, valid for any antisymmetric matrix
\begin{equation}\label{Fid}
    F\halfspace\dualF = \tfrac14 \tr(F\halfspace\dualF) \halfspace\unit
\end{equation} 
where $\unit$ is the unit matrix.   Note that $X_2$ is duality-invariant and is essentially the (symmetric) stress-energy tensor which is given by
\begin{equation}\label{T}
   T = -2\Fscr \bar\Fscr = -\tfrac12 (F^2 + \dualF^2) \ .
\end{equation}
It follows that its trace vanishes identically. Therefore, in the standard Maxwell theory we have that $X_1 = 4L$. However, we can have a consistent duality-symmetric formulation if we treat $F_{ab}$ and $G_{ab} := \dualF_{ab}$ as independent quantities. They must then also have independent potentials leading to the introduction of a magnetic potential $C_a$ such that $G_{ab} = -2C_{[a,b]}$. Then, formally, $\tr X_2$ does not vanish as long as the identification $G_{ab} = \dualF_{ab}$ is not imposed. Referring to \eqref{FF} we will therefore used scaled versions of $\tr T$ as duality-symmetric Lagrangians in the following. In addition to the duality transformation \eqref{duality}, a corresponding transformation at the potential level is also needed for expressions involving the potentials. It has the analogous form
\begin{equation}\label{pduality}
 \begin{split}
        A_a &\rightarrow
  \phantom{-}A_a \halfspace\cos\alpha + C_a	 \halfspace\sin\alpha \\[1pt]
   C_a &\rightarrow
            -A_a \halfspace\sin\alpha + C_a \halfspace\cos\alpha
 \end{split}
\end{equation}
We emphasize that due to the additional degrees of freedom represented by $C_a$ and its corresponding field strength $G_{ab}$, the duality-symmetric formulation of Maxwell theory considered below is an extension of standard Maxwell theory, which becomes equivalent to the standard theory only after imposing the duality constraint 
\begin{align}\label{eq:dc}  
G_{ab} = \dualF_{ab} \ .
\end{align} 

Based on the above discussion (cf.~also \cite{Cameron&Barnett:2012,Bliokh_etal:2013}) we define the duality-symmetric Lagrangian
\begin{equation}\label{rds_lag}
   \Lscr = -\tfrac18 (F_{ab} F^{ab} + G_{ab} G^{ab})
       = -\tfrac12\kappa^{abcd} (A_{a,b}A_{c,d} + C_{a,b}C_{c,d}) \ .
\end{equation}
Using the notations $M^a = \Esf_{\text A}^a(\Lscr)$ and $N^a = \Esf_{\text C}^a(\Lscr)$, the Euler-Lagrange expressions are
\begin{equation}
 \begin{split}
   M_a = \tfrac12 F^{b}{}_{a,b}
       = \tfrac12 \square A_a - \tfrac12 (\div A)_{,a} \ ,\\[3pt]
   N_a = \tfrac12 G^{b}{}_{a,b}
       = \tfrac12 \square C_a - \tfrac12 (\div C)_{,a} \ .
 \end{split}
\end{equation}

Using the complex vector potential $\Ascr_a = A_a + i\halfspace C_a$ and the complex field strength $\Fscr_{ab} = -\Ascr_{[a,b]}$, the duality-symmetric Lagrangian \eqref{rds_lag} takes the form (\emph{cf.~}\cite{Bliokh_etal:2013})
\begin{equation}\label{ds_lag}
   \Lscr = -\tfrac12 \Fscr_{ab} \barFscr^{ab} 
         = -\tfrac12\kappa^{abcd} \Ascr_{a,b} \, \barAscr_{c,d}
\end{equation}
 where the bar denotes complex conjugation.
In \cite{Bliokh_etal:2013}, the real and imaginary parts of $\Ascr_a$ were taken as the basic field variables. Here, we will instead use $\Ascr_a$ and $\bar\Ascr_a$ as the basic fields. Since they represent linearly independent combinations of $A_a$ and $C_a$, they can be used as alternative field variables. They carry the same number of degrees of freedom (4 each) and are treated as formally independent. 
Defining $\Mscr^a = \Esf^a_\Ascr(\Lscr)$,  where $\Esf_\Xscr$ is the Euler operator for a variable $\Xscr$, the Euler-Lagrange expressions are given by
%
\begin{equation}
 \begin{split}
   \Mscr^a &= -\frac{\pd}{\pd x^b} \frac{\pd\Lscr}{\pd\Ascr_{a,b}}
            = \tfrac12\barFscr^{ba}{}_{,b}
 \end{split}                  
\end{equation}
and $\Esf_{\!\bar\Ascr}^a(\Lscr) = \bar \Mscr^a$.
Expressed in terms of the potentials they become
\begin{equation}
    \Mscr_a = \tfrac14 \square \barAscr_a -\tfrac14 (\div\barAscr)_{,a} \,.
\end{equation}

\begin{table}[!t]\label{table}
\begin{adjustwidth}{-1in}{-1in}  
\begin{center}
\caption{Summary of Formulas.}\label{table}
\renewcommand{\arraystretch}{1.5}
\resizebox{1.2\textwidth}{!}{
\begin{tabular}{lccc} 
\hline
quantity & symbol & real formulation & complex formulation \\ 
\hline
duality-symmetric Lagrangian & $\Lscr$ & \(\displaystyle -\tfrac18 (F_{ab} F^{ab} + G_{ab} G^{ab}) \) &  \(\displaystyle-\tfrac12 \Fscr_{ab} \barFscr^{ab} \) \\
zilch symmetry & $v_{ab}$ &  \(\displaystyle -2\, G_{c(a,b)}\, \tfrac{\pd}{\pd A_c} + 2\, F_{c(a,b)}\, \tfrac{\pd}{\pd C_c} \) & \(\displaystyle-4i \Fscr_{c(a,b)} \, \tfrac{\pd}{\pd \Ascr_c} + c.c.\) \\
Lagrangian transformation & $\pr v_{ab}(\Lscr)$ & $(-\tfrac12 \delta_{(a{}}{}^c \, G^{de}{}_{,b)} F_{de}  +\tfrac12 \delta_{(a{}}{}^c \, F^{de}{}_{,b)} G_{de})_{,c}$ & \(\displaystyle (-i \delta_{(a{}}{}^c \, \Fscr^{de}{}_{,b)} \barFscr_{de} + \cc)_{,c} \) \\
zilch tensor & $Z_{abc}$ & \(\displaystyle \dualF^d{}_{(a} F_{b)d,c} - F^d{}_{(a} \dualF_{b)d,c}\) & \(\displaystyle2i \bar\Fscr^d{}_{(a} \Fscr_{b)d,c} + \cc \) \\
equivalent zilch tensor\tablefootnote{The equivalence of $Z_{abc}$ and $Z'{}_{abc}$ is shown in \eqref{equivalence_equation}. See also remark \ref{rem:ZZ'}.} & $Z'{}_{abc}$ & \(\displaystyle- F_c{}^{d} \dualF_{d(a,b)} + \dualF_c{}^{d} F_{d(a,b)} \) & \(\displaystyle 2i \bar\Fscr_{d(a,b)} \Fscr^{d}{}_c + c.c. \) \\[2pt]
\hline
\end{tabular}}
 \end{center}
\end{adjustwidth}
\end{table}

\section{Real variational formulation of zilch symmetry}
\label{realdual}
%

%
%
%
%
We now introduce, using the anticommutator notation, cf.\  footnote \ref{foot:anti}, the following expression\footnote{We use the notation $\Zscr$ for the zilch tensor in extended duality-symmetric formulation, in contrast to $Z$ for it in reduced theory where the duality constraint is imposed.}, 
\begin{equation}\label{RDSzilch}
   \Zscr_c = \tfrac12 \{ G, \pd_c F \} - \tfrac12 \{ F, \pd_c G \} \ .
\end{equation}
which upon imposing the duality constraint \eqref{eq:dc} becomes the zilch tensor \eqref{Z3}. 
In component form, \eqref{RDSzilch} is 
\begin{equation}
   \Zscr_{abc} = G^d{}_{(a} F_{b)d,c} - F^d{}_{(a} G_{b)d,c} \ .
\end{equation}
Its divergence is then given by
\begin{equation}\label{divRDSz}
  \Zscr^c{}_{,c} = \tfrac12 \{ G, \square F \}
             - \tfrac12 \{ F, \square G \}
\end{equation}
and in components by
\begin{equation}\label{divRDSzcomp}
   \Zscr_{ab}{}^c{}_{,c} = G^d{}_{(a} \,\square F_{b)d}
                      -F^d{}_{(a} \,\square \halfspace G_{b)d} \,.
\end{equation}

%
Using the relations
\begin{equation}
\square F_{ab} = -4 M_{[a,b]} \ ,\qquad 
\square G_{ab} = -4 N_{[a,b]} \ ,
\end{equation}
we can write the divergence of the zilch tensor in the form
\begin{equation}
    \Zscr_{ab}{}^c{}_{,c} = -4 G_{(a}{}^d \kappa_{b)d}{}^{ef} M_{e,f}
                        +4 F_{(a}{}^d \kappa_{b)d}{}^{ef} N_{e,f} \ .
\end{equation}
%
%
We note that this equation is not in the desired characteristic form since it is not proportional to the Euler-Lagrange expressions but rather to the derivatives of those expressions. However, we can deal with this issue by performing partial integrations leading to
\begin{align}\label{RDSZchar}
\Zscr_{ab}{}^c{}_{,c}
&= -(4 G_{(a}{}^e \kappa_{b)e}{}^{cd} M_c)_{,d} - 2\, G_{c(a,b)} M^c
+(4 F_{(a}{}^e \kappa_{b)e}{}^{cd} N_c)_{,d} + 2\, F_{c(a,b)} N^c \,.
\end{align}
%
This relation can be expressed in the characteristic form
\begin{equation}\label{ZtGF}
    \widetilde{\Zscr}_{ab}{}^c{}_{,c} = - 2\, G_{c(a,b)} M^c  + 2\, F_{c(a,b)} N^c
\end{equation}
for the modified zilch tensor
\begin{equation}\label{RDSZmod}
    \tilde{\Zscr}_{ab}{}^c = \Zscr_{ab}{}^c - 4 G_{(a}{}^e \kappa_{b)e}{}^{cd} M_d + 4 F_{(a}{}^e \kappa_{b)e}{}^{cd} N_d \ .
\end{equation}
which is equal to the zilch tensor $\Zscr_{ab}{}^c$ on-shell in the duality-symmetric theory. The difference ${\Zscr}_{ab}{}^c - \tilde {\Zscr}_{ab}{}^c$ is itself trivially conserved being zero on-shell (referred to as triviality of the first kind in \cite{Olver:1993}). From the general (if and only if) form of Noether's theorem (Olver \cite[Theorem 5.58]{Olver:1993}) it follows from the relation \eqref{ZtGF} that the conservation of the zilch tensor comes from a variational symmetry of the duality-symmetric Lagrangian.

Turning now to the characterization of the explicit form of the corresponding Noether symmetry, we can identify the coefficients of $M^c$ and $N^c$ in \eqref{ZtGF} with the components of the characteristic as
\begin{equation} \label{eq:QP}
Q_{abc} =  -2\, G_{c(a,b)}  \ ,\qquad
P_{abc} =  2\, F_{c(a,b)}  \ ,
\end{equation}
leading to the Noether symmetry candidate
\begin{align}\label{RDSsymm_vec}
v_{ab} ={}& Q_{abc}\, \frac{\pd}{\pd A_c} + P_{abc}\, \frac{\pd}{\pd C_c} \nonumber\\
={}& -2\, G_{c(a,b)}\, \frac{\pd}{\pd A_c} + 2\, F_{c(a,b)}\, \frac{\pd}{\pd C_c} \ ,
\end{align}
which is invariant with respect to the duality transformation \eqref{pduality}. The infinitesimal transformations of potential fields under \eqref{RDSsymm_vec} are
\begin{subequations}\label{eq:AC-sym}
\begin{align}
    A_c &\to A_c -  \zeta^{ab}\, G_{ca,b} \\
    C_c &\to C_c +  \zeta^{ab}\, F_{ca,b}
\end{align}
\end{subequations}
from which, using Bianchi identities, we have
\begin{subequations}\label{eq:FG-sym}
\begin{align}
    F_{cd} &\to F_{cd} +  \zeta^{ab}\, G_{cd,ab} \\
    G_{cd} &\to G_{cd} -  \zeta^{ab}\, F_{cd,ab}
\end{align}
\end{subequations}
for an infinitesimal symmetric matrix $\zeta^{ab}$ of parameters. 
The transformation generated by \eqref{eq:AC-sym}, \eqref{eq:FG-sym} is the same as in \cite[Eq.~(6.20),~(6.21)]{Cameron&Barnett:2012}. See \cite[Sect.~6.3]{Cameron&Barnett:2012} for discussion and references. These symmetries arise from the modified zilch tensor $\tilde{\Zscr}_{ab}{}^c$ in the duality-symmetric theory, defined in \eqref{RDSZmod}.  As we shall demonstrate in the next section, the action of the symmetry \eqref{eq:AC-sym}, \eqref{eq:FG-sym} on the duality-symmetric Lagrangian is a total divergence. As emphasized above for Eq.~\eqref{varsymm} this must hold off-shell in order to qualify as a variational symmetry in accordance with the Noether's theorem.

We show that the duality-symmetric Lagrangian \eqref{rds_lag} is invariant under the symmetry generated by the generalized vector field \eqref{RDSsymm_vec} and the conserved current derived from that symmetry is indeed the zilch current. The prolongation of \eqref{RDSsymm_vec} acting on the duality symmetric Lagrangian gives
\begin{align}
\pr v_{ab}(\Lscr) 
&{}= Q_{abc,d} \,\frac{\pd \Lscr}{\pd A_{c,d}} + P_{abc,d}\, \frac{\pd \Lscr}{\pd C_{c,d}} = \Uscr_{ab}{}^c{}_{,c} \label{RDSprvLDS}
\end{align}
where
\begin{align}
    \Uscr_{ab}{}^c = -\tfrac12\, \delta_{(a{}}{}^c \, G^{de}{}_{,b)} F_{de}  +\tfrac12 \delta_{(a{}}{}^c \, F^{de}{}_{,b)} G_{de}
\end{align}
and we have used the identities 
\begin{align}
G^{cd}\,F_{bd,c,a}=\tfrac12\,G^{cd}\,F_{cd,a,b} \quad \text{and} \quad F^{cd}\,G_{bd,c,a}=\tfrac12\,F^{cd}\,G_{cd,a,b}\,,
\end{align} 
which are consequences of the Bianchi identities $F_{[ab,c]} = G_{[ab,c]}=0$. The fact that \eqref{RDSprvLDS} is a total divergence, verifies that $v_{ab}$ is a symmetry which gives rise to a conservation law from its action on the duality-symmetric Lagrangian. We stress that equation \eqref{RDSprvLDS} is valid off-shell in the duality-symmetric theory, as required by the Noether theory. We note that Cameron and Barnett in \cite{Cameron&Barnett:2012} considered only symmetries valid on-shell.

The Noether current generated by $v_{ab}$ is
\begin{align}\label{rzfromsymm}
\Zscr'{}_{ab}{}^c ={}& -Q_{abd}\,\frac{\pd \Lscr}{\pd A_{d,c}} - P_{abd}\, \frac{\pd \Lscr}{\pd C_{d,c}} + \Uscr_{ab}{}^c \nonumber\\
={}&  - F^{cd} G_{d(a,b)} + G^{cd} F_{d(a,b)} + \Uscr_{ab}{}^c \ ,
\end{align}
where 
$Q_{abc}, P_{abc}$ are given in \eqref{eq:QP}. The current \eqref{rzfromsymm} is equivalent to the modified zilch tensor \eqref{RDSZmod}, which can be seen from the fact that their difference
\begin{align}\label{zilchdiff}
    \Zscr'{}_{ab}{}^c - \tilde{\Zscr}_{ab}{}^c = 2 \,( F_{(a}{}^{[c} G_{b)}{}^{d]} - \delta_{(a}{}^{[c} G_{b)e} F^{d]e} + \delta_{(a}{}^{[c} F_{b)e} G^{d]e} )_{,d}
\end{align}
is obviously a trivial current, i.e. one whose divergence (with respect to index $c$) vanishes identically (triviality of the second kind). Finally, the difference of the currents $\Zscr'{}_{ab}{}^c$ and $\Zscr_{ab}{}^c$ is given by 
\begin{align}\label{equivalence_equation}
    \Zscr'{}_{ab}{}^c - \Zscr_{ab}{}^c =&{}  - 4 G_{(a}{}^e \kappa_{b)e}{}^{cd} M_d + 4 F_{(a}{}^e \kappa_{b)e}{}^{cd} N_d \nonumber\\
    &{}+ 2 \,( F_{(a}{}^{[c} G_{b)}{}^{d]} - \delta_{(a}{}^{[c} G_{b)e} F^{d]e} + \delta_{(a}{}^{[c} F_{b)e} G^{d]e} )_{,d} \ ,
\end{align}
which shows clearly their equivalence involving both kinds of trivialities  mentioned above.

\begin{remark} \label{rem:ZZ'}
Imposing the duality constraint $G_{ab}= \dualF_{ab}$, the last term $\Uscr_{ab}{}^c$ in the current \eqref{rzfromsymm} vanishes, as can be seen by a short calculation.
The remaining terms in \eqref{rzfromsymm} result in
\begin{align} \label{eq:3.30}
- F^{cd} \dualF_{d(a,b)} + \dualF^{cd} F_{d(a,b)}\, ,
\end{align}
 which we shall denote by  $Z'{}_{ab}{}^c$ (in contrast to its nontrivial extension $\Zscr'{}_{ab}{}^c$ in duality-symmetric formulation). It is a form of the zilch tensor introduced by Anco and Pohjanpelto\footnote{Our definition of $Z'{}_{ab}{}^c$ is related to Anco and Pohjanpelto's by an overall minus sign and reordering the indices to agree with our convention for the conservation law $Z'{}_{ab}{}^c{}_{,c} \weak 0$.} in \cite{Anco&Pohjanpelto:2001}. 
%
The two forms $Z_{ab}{}^c$ and $Z'_{ab}{}^c$ of the zilch tensor are equivalent in the sense that they contain the same information (off-shell) (cf.~ \cite[p.143]{Penrose&Rindler_1:1986}). This is expressed by the relations
\begin{equation}
   Z'{}_{abc} = Z_{c(ab)}
\end{equation}
and its converse
\begin{equation}
   Z_{abc} = -2 Z'{}_{abc} + 3 Z'{}_{(abc)} \, .
\end{equation}
\end{remark}

\section{Complex variational formulation of zilch symmetry}
\label{complexdual}
%


%
%
The complex duality-symmetric formulation of zilch conservation parallels that of the real formulation in Sect.~\ref{realdual}. Rewriting the zilch expression \eqref{RDSzilch} in terms of $\Fscr^a{}_b$ gives
\begin{equation}\label{sdzilch}
   \Zscr_c = i \{\bar\Fscr, \pd_c \Fscr \} + \cc
\end{equation}
or in component form
\begin{equation}
    \Zscr_{abc} = 2i \bar\Fscr^d{}_{(a} \Fscr_{b)d,c} + \cc 
\end{equation}
where ``\cc" stands for the complex conjugate of the preceding expression after the equality sign.
Its divergence is then given by
\begin{equation}\label{divsdzilch}
   \Zscr^c{}_{,c} = i\{\bar\Fscr, \square\Fscr\} + \cc 
\end{equation}
or in components
\begin{equation}\label{divsdzcomp}
   \Zscr_{ab}{}^c{}_{,c} = 2i\bar\Fscr^d{}_{(a} \,\square\Fscr_{b)d} + \cc
\end{equation}
%
Using the relation
\begin{equation}
    \square \Fscr_{ab} = -4 \bar\Mscr_{[a,b]}  ,
\end{equation}
and its complex conjugate, we express \eqref{divsdzcomp} in the form  (cf.~ \eqref{RDSZchar})
\begin{equation}
   \Zscr_{ab}{}^c{}_{,c}
     = -4i\Fscr_a{}^d \,\Mscr_{[d,b]}
       -4i\Mscr_{[a,d]} \, \Fscr^d{}_b
       + \cc \ .
\end{equation}
Performing partial integration now leads to a conservation law in the characteristic form
\begin{equation}\label{Zchar}
  \tilde\Zscr_{ab}{}^c{}_{,c}
    = \mathcal Q_{ab}{}^c \Mscr_c + \cc
\end{equation}
where
\begin{equation}
    \tilde\Zscr_{abc} = \Zscr_{abc}
                       -8i\Fscr_{e(a} \kappa_{b)}{}^{ecd} \Mscr_d
\end{equation}
and
\begin{equation}
   \mathcal Q_{abc} = -4i \Fscr_{c(a,b)}
\end{equation}
is the characteristic.
The corresponding Noether symmetry candidate is then given by
\begin{equation}\label{symm_vec}
   v_{ab} = \mathcal Q_{abc} \frac{\pd}{\pd \Ascr_c} + \bar{\mathcal Q}_{abc} \frac{\pd}{\pd \barAscr_c} \ ,
\end{equation}
which is manifestly duality invariant.

We show that the duality-symmetric Lagrangian \eqref{ds_lag} is invariant under the symmetry generated by the generalized vector field \eqref{symm_vec} and the conserved current derived from that symmetry is indeed the zilch current. The prolongation of \eqref{symm_vec} acting on duality-symmetric Lagrangian gives
\begin{align}
    \pr v_{ab}(\Lscr) 
    &{}= \mathcal Q_{abc,d} \,\frac{\pd \Lscr}{\pd \Ascr_{c,d}} + \bar{\mathcal Q}_{abc,d}\, \frac{\pd \Lscr}{\pd \barAscr_{c,d}} = \Uscr_{ab}{}^c{}_{,c} \ , \label{prvLDS}
\end{align}
where
\begin{align}
    \Uscr_{ab}{}^c = -i \delta_{(a{}}{}^c \, \Fscr^{de}{}_{,b)} \barFscr_{de}                 + \cc
\end{align}
and we have used the identity $\barFscr^{cd}\,\Fscr_{bd,c,a}=\tfrac12\,\barFscr^{cd}\,\Fscr_{cd,a,b}$, which is the consequence of the Bianchi identity $\Fscr_{[ab,c]}=0$. To verify that this symmetry results in the zilch tensor as its conserved current, we note that the conserved current is given by
\begin{align}\label{zfromsymm}
    \Zscr'{}_{ab}{}^c ={}& -\mathcal Q_{abd}\,\frac{\pd \Lscr}{\pd \Ascr_{d,c}} - \bar{\mathcal Q}_{abd}\, \frac{\pd \Lscr}{\pd \barAscr_{d,c}} + \Uscr_{ab}{}^c \,.
\end{align}
 To convert this to a current for solutions of the Maxwell equations, we need to impose the duality constraint \eqref{eq:dc}, which breaks the independence of $\Fscr_{ab}$ and $\bar\Fscr_{ab}$ that has been assumed in this section. One then finds that $\Uscr_{ab}{}^c$ vanishes and the remaining terms in \eqref{zfromsymm} result in the equivalent zilch tensor (cf. remark \ref{rem:ZZ'})
\begin{align}
Z'_{ab}{}^c = 2i \bar\Fscr_{d(a,b)} \Fscr^{dc}
                 -2i \Fscr_{d(a,b)} \bar\Fscr^{dc} \, .
\end{align}

%
\section{Concluding remarks}\label{sec:concl}
We have discussed how the conservation of the zilch tensor is related to a variational symmetry in accordance with the Noether theory. The fact that the zilch tensor is conserved has been known since its discovery by Lipkin \cite{Lipkin:1964}. In this paper, we have derived the variational symmetry generator for a version of the zilch tensor in the duality-symmetric theory. The existence of this symmetry generator is guaranteed by Noether's theorem. The proof of its variational nature in accordance with \eqref{conslaw1} is given by \eqref{RDSprvLDS}.
We note that the generator itself was given by Cameron and Barnett \cite{Cameron&Barnett:2012}, however without proof of its variational nature.

To achieve the direct correspondence between the symmetry generator and the conservation law in the characteristic form \eqref{conslaw2}, it was necessary to augment the zilch tensor by a trivial addition which vanishes on-shell, see \eqref{RDSZmod}. This feature is a reflection of the fact that symmetries and corresponding conservation laws are in general only determined as equivalence classes of symmetries and conservation laws. Members of an equivalence class of symmetries or conservation laws can differ by trivial symmetries or trivial conservation laws respectively \cite{Olver:1993}. 
We should note that the considerations here do not rule out the possibility that the standard Lagrangian could be invariant under a symmetry that generates the zilch tensor as a Noether current. In fact, a careful examination reveals that an expression closely related to \eqref{RDSsymm_vec} generates a variational symmetry of the standard Lagrangian. We will explore this subject in a future work. 

In section \ref{complexdual}, we have derived the zilch symmetry and its associated conservation law in a complex duality-symmetric formulation of Maxwell theory. Here the complex set-up is only used in an ad-hoc way. However, the spacetime has an intrinsic complex structure which can be revealed in a geometric (Clifford) algebra approach to the spacetime (a comprehensive review of this subject and its application in electromagnetism can be found in \cite{DRESSEL20151}). While the conservation laws of helicity, and the spin and orbital parts of angular momentum were studied there, it can be interesting to perform a similar analysis for the zilch tensor and its conservation law in that framework. We expect that a complete analysis along the lines of this paper, based on the jet space formalism and using spacetime algebra, is possible and may shed  new light on the complex formulation applied in this paper.


\subsection*{Acknowledgements} S.A. thanks Professor Golshani for his support and encouragement. K.R.  thanks the Albert Einstein Institute, Potsdam, where a substantial part of this work was done, for hospitality and support. L.A. thanks the Royal Institute of Technology and University of Stockholm for hospitality and support during work on this paper. Part of this work was done while L.A. and K.R. were in residence at Institut Mittag-Leffler in Djursholm, Sweden during the fall of 2019, supported by the Swedish Research Council under grant no.~2016-06596. We also thank the referee for valuable comments on the previous version of the paper.

\appendix
\section{Different forms of the zilch tensor and their equivalence}\label{app:alternative_forms}
For ease of reference, we collect here the different forms of the zilch tensor used in the paper. We show these forms are all equivalent currents in the sense of section \ref{jet_intro}. In section \ref{sec:zilch} we discussed the three versions of the zilch tensor introduced by Kibble \cite{Kibble:1965} in standard Maxwell theory,
\begin{align}
	Z_{abc} 
		&= \dualF_{ad} F^d{}_{b,c}-F_{ad}\halfspace\dualF^d{}_{b,c} \\
		&= 2\halfspace\dualF^d{}_{(a} F_{b)d,c}
		-\tfrac12 g_{ab} \,\dualF^d{}_e \halfspace F^e{}_{d,c} \\
		&= \dualF^d{}_{(a} F_{b)d,c} - F^d{}_{(a} \dualF_{b)d,c} \ .
\end{align}
Applying the symmetry transformation associated to the conservation of the zilch tensor according to Noether's theorem, we encountered in \eqref{eq:3.30} another version of the zilch tensor, \emph{cf}.~\cite{Anco&Pohjanpelto:2001}, 
\begin{align}\label{zilch_Anco}
Z'_{ab}{}^c = \dualF^{cd} F_{d(a,b)} - F^{cd} \dualF_{d(a,b)}  \ .
\end{align}
The original definition of zilch tensor was made by Lipkin \cite{Lipkin:1964} in a form that can be written more compactly (using the dual field) as
\begin{align}
 \dualF^{cd} F_{d(a,b)} - F^{cd} \dualF_{d(a,b)} + 2\,(\dualF_{(a}{}^{[c}\,F_{b)}{}^{d]})_{,d}  \ .
\end{align}
in which the third term is a trivial current and the first two terms coincide with $Z'_{abc}$ in \eqref{zilch_Anco}.

\section{$1+3$ decomposition of the zilch tensor}\label{app:decomposition}
For reference we list the full time-space decomposition of the components of Kibble's form \eqref{ZKibble} of the zilch tensor, both before and after imposing the field equations.
\begin{align}
Z_{000} &= u^a u^b u^c Z_{abc} = \tb E \cdot \dot{\tb{B}} - \tb B \cdot \dot{\tb{E}} \weak - \tb E \cdot (\nabla \times \tb E) - \tb B \cdot (\nabla \times \tb B) 
\\
Z_{00i} &= u^a u^b e^c_i Z_{abc} =  E^j B_{j,i} - B^j E_{j,i} \weak  Z_{i00} + \left[(\tb E \cdot \nabla)\tb B - (\tb B \cdot \nabla)\tb E\right]_i 
\\
Z_{i00} &= e^a_i u^b u^c Z_{abc} = (\tb E \times \dot{\tb{E}} + \tb B \times \dot{\tb{B}})_i \weak \left[\tb E \times (\nabla\times \tb B) - \tb B \times (\nabla\times \tb E)\right]_i
\\
Z_{i0j} &= Z_{0ij} = u^a e^b_i e^c_j Z_{abc} = \epsilon_{ikl} (E^k E^l{}_{,j} +  B^k B^l{}_{,j}) \nonumber\\
&= -\delta_{ij}(\tb E \cdot (\nabla \times \tb E) + \tb B \cdot (\nabla \times \tb B)) + (\nabla\times \tb{E})_i E_j + (\nabla\times\tb{B})_i B_j \nonumber\\
&\quad -\epsilon_{i}{}^{kl} (E_{j,k}E_l + B_{j,k}B_l) \nonumber\\
&\weak  Z_{ij0} - E_i (\nabla\times \tb{E})_j - B_i (\nabla\times \tb{B})_j -\epsilon_{i}{}^{kl} (E_{j,k}E_l + B_{j,k}B_l)
\\
Z_{ij0} &= e^a_i e^b_j u^c Z_{abc} = \delta_{ij} (\tb E \cdot \dot{\tb{B}} - \tb B \cdot \dot{\tb{E}})  + 2 \dot{E}_{(i} B_{j)} - 2 \dot{B}_{(i} E_{j)} \nonumber\\
&\weak  \delta_{ij} Z_{000} + 2 (\nabla\times\tb{B})_{(i} B_{j)}
  + 2 (\nabla\times \tb{E})_{(i} E_{j)} \\
Z_{ijk} &= e^a_i e^b_j e^c_k Z_{abc} = \delta_{ij} (E^l B_{l,k} - B^l E_{l,k}) - 2 E_{(i}B_{j),k} + 2 B_{(i}E_{j),k} \nonumber\\
&= \delta_{ij} Z_{00k} - 2 E_{(i}B_{j),k} + 2 B_{(i}E_{j),k}
\end{align} 

\singlespacing
\bibliographystyle{unsrt}
\bibliography{references}

\end{document}